\documentstyle[12pt,oldlfont]{article}

\newlength{\aivwidth}   
\newlength{\tmpwidth}   

\newcommand{\be}{\begin{equation}}
\newcommand{\ee}{\end{equation}}
\newcommand{\bea}{\begin{eqnarray}}
\newcommand{\eea}{\end{eqnarray}}
\newcommand{\ba}{\begin{array}}
\newcommand{\ea}{\end{array}}
\newcommand{\eref}[1]{(\ref{#1})}
\newcommand{\nn}{\nonumber\\}

\renewcommand{\theequation}{\thesection.\arabic{equation}}
\newcounter{saveeqn}
\newcommand{\alpheqn}{\setcounter{saveeqn}{\value{equation}}%
   \addtocounter{saveeqn}{1} \setcounter{equation}{0}%
   \renewcommand{\theequation}%
   {{\thesection.\arabic{saveeqn}\alph{equation}}}}
\newcommand{\reseteqn}{\setcounter{equation}{\value{saveeqn}}%
   \renewcommand{\theequation}{\thesection.\arabic{equation}}}
\newcommand{\alphtab}{\setcounter{saveeqn}{\value{table}}%
   \addtocounter{saveeqn}{1} \setcounter{table}{0}%
   \renewcommand{\thetable}%
   {{\arabic{saveeqn}\alph{table}}}}
\newcommand{\resettab}{\setcounter{table}{\value{saveeqn}}%
   \renewcommand{\thetable}{\arabic{table}}}

\newcommand{\eqnew}{\setcounter{equation}{0}}
\newcommand{\phrd}[1]{Phys.\ Rev.\ {\bf D#1}}

\newcommand{\phrl}[1]{Phys.\ Rev.\ Lett.\ {\bf #1}}
\newcommand{\nphb}[1]{Nucl.\ Phys.\ {\bf B#1}}

\newcommand{\phlb}[1]{Phys.\ Lett.\ {\bf B#1}}

\newcommand{\zphc}[1]{Z.\ Phys.\ {\bf C#1}}
\newcommand{\aph}[1]{Ann.\ Phys.\ (N.Y.) {\bf #1}}

\newcommand{\suu}{$\rm SU(2)\times U(1)$}
\newcommand{\uem}{$\rm U(1)_{\rm em}$}
\newcommand{\rx}{$\rm R_\xi$}
\newcommand{\lag}{{\cal L}}
\newcommand{\ep}{\epsilon}
\newcommand{\tr}{\,\mbox{tr}\,}
\newcommand{\vp}{\phi}
\newcommand{\lsm}{\mbox{$\lag_{SM}$}}
\newcommand{\lgnlsm}{\mbox{$\lag_{GNLSM}$}}
\newcommand{\lw}{\mbox{$\lag_{W}$}}
\newcommand{\lwp}{\mbox{$\lag_{W\Phi}$}}
\newcommand{\lbp}{\mbox{$\lag_{B\Phi}$}}
\newcommand{\lww}{\mbox{$\lag_{WW}$}}
\newcommand{\lbb}{\mbox{$\lag_{BB}$}}
\newcommand{\ld}{\mbox{$\lag_{D}$}}
\newcommand{\lwi}{\mbox{$\lag_{W}^\prime$}}
\newcommand{\lwpi}{\mbox{$\lag_{W\Phi}^\prime$}}
\newcommand{\lddi}{\mbox{$\lag_{DD1}^\prime$}}
\newcommand{\lddii}{\mbox{$\lag_{DD2}^\prime$}}
\renewcommand{\sp}{@{~~~}}
\renewcommand{\ss}{}
\newcommand{\yes}{$\times$}
\newcommand{\m}{$-$}
\newcommand{\fb}[1]{\fbox{#1}}
\newcommand{\vl}{V_L}
\newcommand{\vt}{V_T}

\newcommand{\M}{{\cal M}}

\begin{document}


\begin{titlepage}
\title{The Equivalence Theorem and Effective Lagrangians}
\author{Carsten Grosse-Knetter\thanks{E-Mail:
knetter@physw.uni-bielefeld.de}
\\and\\Ingolf Kuss\thanks{E-Mail:
kuss@physw.uni-bielefeld.de}\\[5mm]
Universit\"at Bielefeld\\Fakult\"at f\"ur Physik\\Postfach 10 01 31\\
33501 Bielefeld\\Germany}
\date{BI-TP 94/10\\hep-ph/9403291\\March 1994}
\maketitle
\thispagestyle{empty}

\begin{abstract}
We point out that
the equivalence theorem, which relates the amplitude for a process
with external longitudinally polarized vector bosons to the amplitude
in which the longitudinal vector bosons are replaced by the
corresponding pseudo-Goldstone bosons, is not valid for effective
Lagrangians. However, a more general formulation of this
theorem also holds for effective interactions.
The generalized theorem can be utilized
to determine the high-energy behaviour of
scattering
processes just by power
counting and to simplify the calculation of the corresponding
amplitudes. We apply this method
to the phenomenologically most interesting
terms describing effective
interactions of the electroweak vector and Higgs bosons in order to
examine their effects on
vector-boson scattering and on vector-boson-pair production in
$f\bar{f}$ annihilation.
The use of the equivalence theorem in the literature is examined.
\end{abstract}
\end{titlepage}


\section{Introduction}
\typeout{Section 1}
The equivalence theorem (ET)
\cite{lequth,chga,gokone,ve} has become a useful
tool for investigating the high-energy behaviour of
scattering processes with external
longitudinally polarized vector bosons, especially
vector-boson scattering,
$V_LV_L\to V_LV_L$, and vector-boson-pair
production in fermion-antifermion annihilation, $f\bar{f}\to V_L V_L$.
This theorem states that the amplitude
for such a process is
equal to the amplitude
(calculated within the \rx -gauge) in which the external
longitudinal vector bosons are replaced by the corresponding
pseudo-Goldstone bosons
(times a phase factor),
apart from corrections which are of order
$M/E$ with $M$ being the vector-boson mass and $E$ being its
energy\footnote{This means that amplitudes which decrease with
increasing energy $E$ cannot be determined by applying the ET.}:
\bea&&
\M(V_{L,1}\ldots V_{L,N_1}A\to V_{L,1}\ldots V_{L,N_2}B)\nn
&&\qquad\qquad=i^{N_1}(-i)^{N_2}
\M(\phi_{1}\ldots\phi_{N_1}A\to \phi_{1}\ldots\phi_{N_2}B)
+O\left(\frac{M}{E}\right)
\label{et}\eea
($V_{L,i}$, $\phi_i$,
$A$ and $B$ denote the longitudinal vector bosons, the
corresponding Goldstone
bosons and
all other in- and outgoing particles,
respectively.) The ET was first
proven (within the electroweak standard model (SM))
in \cite{chga}. A simpler proof was
given in \cite{gokone}.

The ET simplifies calculations of $S$-matrix elements, because
on the r.h.s.\ of \eref{et} no gauge cancellations occur.
E.g., if one wants to calculate the amplitude for a process
$V_LV_L\to V_LV_L$ within the SM without using the ET,
one has to consider (in the
tree approximation) a diagram with a four-vector
contact interaction, vector
exchange diagrams and scalar exchange diagrams.
The single Feynman diagrams grow with increasing total energy $E$ as
$E^4$ (contact and vector-exchange)
or $E^2$ (scalar exchange) for $E\gg M_H$.
However, when summing up the diagrams
all terms which grow with positive powers of $E$ cancel
and the resulting amplitude is proportional to $E^0$.
Thus, to calculate
the $O(E^0)$-part of the amplitude one also
has to consider the nonleading contributions of the single diagrams.
If one
applies instead the ET one has
to calculate $\phi\phi\to\phi\phi$, viz.\ contact, Higgs-exchange
and vector-exchange diagrams with four external Goldstone bosons.
All these diagrams are at most $O(E^0)$ and thus no cancellations occur.
Therefore it is sufficient to consider only the leading part of the
single diagrams.

During the last few
years, vector-boson scattering and vector-boson-pair
production have been investigated within effective electroweak
theories which
contain nonstandard interactions among the vector bosons
and nonstandard interactions between vectors and scalars
in order to investigate the effects of these theories on
future experiments. When considering
an effective Lagrangian one assumes
that there exists an (unknown)
renormalizable theory (``new physics'') which
involves in addition to the known
particles also heavy particles. At an
energy scale much lower than the mass of the heavy particles
these can be removed from the theory by expressing
their effects through effective (nonrenormalizable)
interactions of the light particles.
Two scenarios how such an effective
theory can be generated have been investigated in the literature:
\begin{enumerate}
\item
If there is a relatively light
Higgs boson, the effective Lagrangian consists of the
SM Lagrangian plus additional
effective interaction terms of higher dimension
which are also gauge invariant
with respect to the electroweak gauge group
\suu\ and in which the scalar
sector is linearly realized like in the
SM Lagrangian \cite{lelora}.
\item
On the other hand one can assume that the Higgs boson
is heavy or even does not exist, i.e. the symmetry breaking
sector is
(at least a part of) the new physics. In this case one
adds effective interaction terms to the Lagrangian of the
gauged nonlinear $\sigma$-model (GNLSM).
The GNLSM is the limit of the SM
for infinite Higgs Mass \cite{apbe}. In this (nonrenormalizable)
gauge theory the scalar sector is
nonlinearly realized. The additional effective terms are also \suu\
gauge invariant with the
scalar sector being nonlinearly realized.
They can be generated by the heavy Higgs sector through
loop effects \cite{apbe} or by other effects of the new physics.
\end{enumerate}

In order to calculate $S$-matrix elements within such an effective
theory, one could be tempted to use the ET. In general, however, this
is not correct because {\em the equivalence theorem is not valid
within effective theories\/}; the proof of
the ET within the SM given
in \cite{chga,gokone,ve} cannot be generalized
to this case. This can easily be seen as follows:
The proof consists of
two steps:
\begin{enumerate}
\item
First, the  following identity for
$S$-matrix elements with external longitudinal vector bosons is
derived:
\bea&&
\M(V_{L,1}\ldots V_{L,N_1}A\to V_{L,1}\ldots V_{L,N_2}B)\nn
&&
=\sum_{M_1=0}^{N_1}\sum_{M_2=0}^{N_2}i^{M_1}(-i)^{M_2}
[\M(\phi_{1}\ldots\phi_{M_1}v_{M_1+1}\ldots v_{N_1}
A\to \phi_{1}\ldots\phi_{M_2}v_{M_2+1}\ldots v_{N_2}B)\nn
&&\qquad\qquad\qquad\qquad\qquad
 +\mbox{permutations of the $\phi$s and $v$s}].
\label{get}\eea
$v$ stands for an external vector boson with its
longitudinal polarization
vector being substituted by
the nonleading part
\be
v^\mu=\epsilon_L^\mu-\frac{P^\mu}{M}
\label{v}\ee
(with $P^\mu$ being the four-momentum of the particle)
which is $O(M/E)$. We will refer to
the relation \eref{get} as the
generalized equivalence theorem (GET). The GET
expresses an $S$-matrix element as a
sum of all amplitudes  that can be constructed by replacing each
longitudinal vector boson either by a Goldstone boson or its
polarization vector $\epsilon_L$ by the nonleading term $v$
\eref{v} (multiplied by appropriate
phase factors).
In distinction to the ET \eref{et}, which only determines the
$O(E^0)$-terms, the GET \eref{get} is a correct relation, in which
no approximation is made.
\item
In the second step,
\eref{et} is proven by showing that all terms on the r.h.s.\
of \eref{get} except for the one with $M_1=N_1$, $M_2=N_2$ are
$O(M/E)$ within the SM.
\end{enumerate}
The proof of the GET \eref{get} is only based on the
BRS invariance and thus on the gauge invariance of the
quantized Lagrangian \cite{chga,gokone,ve}.
It is not affected by adding
effective interaction terms to the theory.
This means that the GET is not only valid for the SM but
also for
any effective gauge theory\footnote{In \cite{chga} this was
shown only for theories with a linearly realized scalar sector.
However the alternative
proof in \cite{gokone} holds for both, linear and
nonlinear models. Recently, a proof analogous to the one in
\cite{chga}
has been done for the nonlinear case \cite{dope}.}.
However, the derivation of the ET
\eref{et} from \eref{get} requires that all amplitudes
calculated within the theory do not increase with increasing
energy, i.e. that they behave at most as  $O(E^0)$ \cite{chga,ve}.
Therefore, the ET is not a consequence of BRS invariance (and thus
gauge invariance) alone, it follows from BRS invariance
{\em and\/}
good high-energy behaviour. The latter requirement is not
fulfilled within effective theories\footnote{In a recent work
\cite{hekuli} it has been claimed that the ET also holds for
effective Lagrangians with additional nonstandard interaction terms.
This is not true; the authors of \cite{hekuli} only observe that the
proof of \eref{get} also holds for effective Lagrangians and they
erroneously conclude that this implies the ET \eref{et}.},
in which
amplitudes in general
increase with increasing energy due to the effects of the
additional interaction terms.

Applying this reasoning to the SM, this means that
a priori the ET only
holds for $E\gg M_H$, because only in this case the amplitudes
behave at most as $O(E^0)$. At energies lower then the Higgs mass
the the amplitudes increase with $E$ \cite{lequth}.
However, it has been shown in \cite{ve} that the ET is also valid
within the SM for $E \ll M_H$. This means that the ET even
holds within the GNLSM, because this is the heavy-Higgs limit of the
SM.

Furthermore we want to remind the reader of the fact that the ET
only states that {\em external\/} longitudinal vector bosons can be
replaced by Goldstone bosons; {\em internal\/} vector lines
still have to be considered. Thus, in general,  {\em
the interactions of longitudinally
polarized vector bosons cannot be described by a
Lagrangian with only scalar fields\/} in contradiction to
the original
formulation of the ET in \cite{lequth}. For example, if one
calculates $\phi\phi\to\phi\phi$ in order to determine the
$O(E^0)$ terms of the amplitude for $V_LV_L\to V_LV_L$
within the SM,
the vector-exchange diagrams cannot be neglected if $E\gg M_H$.

In this article we will show that the GET \eref{get},
which occurs as a byproduct of the proof of the ET
within the SM and which also
holds for effective Lagrangians, is a very usefool tool to determine
the high-energy behaviour of
scattering processes calculated within
effective theories. Our analysis will be
carried out within
tree approximation\footnote{It should be noted that
one has to consider correction factors
stemming from the renormalization of the external
lines if one applies the ET beyond the tree-level
\cite{yayu}.}
which is no severe
restriction since, due to the smallness of the effective
terms, phenomenological
investigations of such terms are usually carried out at the
tree level. The GET \eref{get} expresses the amplitude
for a process with longitudinal vector bosons by a sum of amplitudes
with external $\phi$s and $v$s.
Like in the SM, on the r.h.s.\ of \eref{get} in general
no cancellations occur, i.e.\ the single Feynman diagrams
that contribute to the r.h.s.\ have the
high-energy behaviour that they are supposed to have by power
counting and the leading terms do not cancel
when summing up the various diagrams.
This means that in order to calculate the
leading term of an $S$-matrix element,
one can determine by simple power counting
which of the diagrams on the r.h.s.\ contribute to the
highest powers of $E$.
Then one only has to calculate the leading contributions to
these diagrams.
The only difference to the SM is that, in general, not just those
diagrams with all external $V_L$s replaced by $\phi$s
contribute to the leading terms,
there may be contributions from some diagrams with $v$s.

Such a power-counting method cannot be
applied if the amplitudes are calculated directly without using
the GET, because even in effective theories
cancellations occur \cite{gago,golare,ku}. For example, if one adds
to the SM Lagrangian a dimension-six quadrupole interaction term,
the additional contributions to the
single Feynman diagrams for $V_LV_L\to V_LV_L$
are by power counting supposed to diverge as
$E^6$ but due to cancellations they are $O(E^2)$
and when summing them up the resulting
amplitude turns out to be
only $O(E^0)$ \cite{golare,ku}. We will show below
that this is obvious if one uses the GET,
because all diagrams on the
r.h.s.\ of \eref{get}
are at most $O(E^0)$ within this special effective theory.
Furthermore we will see that this quadrupole interaction
is an example for the invalidity of the ET within
effective theories,
because the quadrupole term yields $O(E^0)$
corrections to the SM amplitude,
while the ET \eref{et} would predict that even those are absent
since this term yields no additional interactions involving
scalar fields. The $O(E^0)$ corrections, however can easily be
calculated by using the GET.
Other dimension-six
interaction terms in general
yield an $O(E^2)$ behaviour
of the amplitudes, although the single diagrams with
external longitudinal
vectors behave as $O(E^4)$ \cite{ku}. This can also
easily be seen from the GET.

In this article we will apply the power-counting method based on
the GET \eref{get} to the phenomenologically most
interesting terms that describe effective interactions of the
electroweak vector and Higgs bosons. We will consider both
linear models with a Higgs boson and nonlinear models without
a Higgs boson. We will analyze the effects of these terms on the
high-energy behaviour of the
processes $VV\to VV$ and $f\bar{f}\to VV$, where the external $V$s
may be longitudinal or transversal.
We will find that our results agree with those obtained by a direct
calculation \cite{ku}.

Although the ET \eref{et}
is not valid for effective Lagrangians, it was
applied in several articles in order to calculate
$S$-matrix elements
within effective theories \cite{dohete,falusi,badava}.
In \cite{dohete}, effective interactions of longitudinal vector
bosons were even described by a Lagrangian with only scalar fields,
i.e.\ even internal vector lines were neglected.
On the basis of our results we will
critically analysize these works. We will find that
indeed in most of these articles the high-energy effects of
several
effective interaction terms are not correctly determined,
however that
the numerical size of the errors is in general small.

This article is organized as follows: In Section~2 we introduce the
effective interaction terms that we will consider. We examine both
linear effective Lagrangians with a Higgs boson and nonlinear ones
without. The validity of the generalized equivalence theorem
\eref{get} within effective
theories is discussed.
We determine which
vertices follow from the effective interaction
terms and the powers of $E$ on that they depend.
In Section~3 we apply the power-counting method based on
the GET \eref{get}
to these effective Lagrangians in order to examine their effects on
the processes $VV\to VV$ and $f\bar{f}\to VV$. We show that the
high-energy behaviour can easily be
determined and that the calculations
become simplified. The results of this analysis are compared with
those obtained by direct calculations.
Phenomenological consequences are discussed. In Section~4
we critically analyze those articles in which the equivalence theorem
has been applied to effective theories. Section~5 contains some
concluding remarks.


\section{The Effective Interaction Terms}
\typeout{Section 2}
\eqnew
In this section we introduce the effective interaction terms that
will be examined later in this article.
We consider both linear
and nonlinear Lagrangians.
We restrict to $P$, $C$ and $CP$ invariant nonstandard interactions.
We discuss these terms only
briefly, because this has been done in more detail elsewhere
\cite{apbe,ku,ruetal,gore,gkkusc}.

We use the following notation:
\alpheqn
\bea
W_\mu&=&\frac{1}{2}W_{\mu i}\tau_i,\label{not1}\\
W_{\mu\nu}&=&\partial_\mu W_\nu-\partial_\nu W_\mu+ig[W_\mu,
W_\nu],\\
B_{\mu\nu}&=&\partial_\mu B_\nu-\partial_\nu B_\mu,\\
\Phi&=&\frac{1}{\sqrt{2}}((v+H){\bf 1}+i\vp_i\tau_i),
\label{phi}\\
D_\mu\Phi&=&\partial_\mu\Phi+igW_\mu\Phi-\frac{i}{2}
g^\prime\Phi\tau_3B_\mu.\label{not2}
\eea
\reseteqn
$W_{\mu i}$ and $B_\mu$ denote the gauge fields, $g$ and $g^\prime$
the coupling contants, $v$ the vacuum expectation value,
$H$ the Higgs field, $\vp_i$ the pseudo-Golstone fields and
$\tau_i$ the Pauli-matrices.

First we consider the case that a Higgs boson exists (with a mass
which is light in comparison to the energy of the scattering process
under consideration).
In this case the effects of new physics can be parametrized by adding
gauge invariant terms of higher (mass) dimension
to the SM Lagrangian
\cite{lelora}.
I.e. one considers an effective Lagrangian
\be
\lag_{eff}=\lag_{SM}+\sum_{n>4}\sum_i\frac{\ep_i}{\Lambda^{n-4}}
\lag_i^{(n)},
\label{leffl}
\ee
where $\lag_{SM}$ is the (dimension-four) SM Lagrangian
\bea
\lag_{SM}&=&-\frac{1}{2}\tr(W_{\mu\nu}W^{\mu\nu})-\frac{1}{4}
B_{\mu\nu}B^{\mu\nu}\nn&&+\frac{1}{2}\tr[(D_\mu\Phi)^\dagger
(D^\mu\Phi)]-\frac{1}{2}\mu^2\tr(\Phi^\dagger\Phi)
-\frac{1}{4}\lambda[\tr(\Phi^\dagger\Phi)]^2\nn&&
+\mbox{fermionic terms,}
\label{lsm}
\eea
$\lag_i^{(n)}$
are the effective interaction terms of dimension $n$,
$\ep_i$ are effective coupling constants and $\Lambda$ is the
scale of the new physics.

The phenomenologically most important
effective terms are those of dimension six,
since terms of higher dimension are suppressed by higher negative
powers of $\Lambda$. The complete list of dimension-six terms is given
in \cite{lelora}. Most of these terms contain fermionic couplings
or affect the gauge-boson propagators. This would yield tree-level
effects on the processes $f\bar{f}\to f\bar{f}$ that have been
measured at LEP~I and other present experiments.
Since no deviations from the SM have been found
(within the experimental accuracy), we know that
the coupling constants of
these terms are very small
\cite{ruetal,gkkusc}. Therefore we restrict ourselves to
those effective
terms, which contain only effective interactions
among the vector bosons and vector--scalar interactions
and thus have no tree-level effects on $f\bar{f}\to
f\bar{f}$ but on the processes
$f\bar{f}\to VV$ and $VV\to VV$, which will be investigated in
future experiments like LEP~II, NLC ($e^+ e^- \to W^+ W^-$)
or LHC ($q\bar{q}\to VV$ and $VV\to VV$ as subprocesses of
$pp \to VVX$). These terms
are \cite{ruetal,gkkusc}:
\alpheqn
\bea
\lag_W&=&-\frac{2}{3}i\tr(W_\mu^{\,\,\,\nu}
W_\nu^{\,\,\,\lambda}W_\lambda^{\,\,\,\mu}),\label{lw}\\
\lag_{W\Phi}&=&i\tr[(D_\mu\Phi)^\dagger W^{\mu\nu}
(D_\nu\Phi)],\label{lwp}\\
\lag_{B\Phi}&=&-\frac{1}{2}i\tr[\tau_3(D_\mu\Phi)^\dagger
(D_\nu\Phi)]B^{\mu\nu},\label{lbp}\\
\lag_{WW}&=&-\frac{1}{8}\tr(\Phi^\dagger\Phi)
\tr(W_{\mu\nu}W^{\mu\nu}),\label{lww}\\
\lag_{BB}&=&-\frac{1}{16}\tr(\Phi^\dagger\Phi)
B_{\mu\nu}B^{\mu\nu},\label{lbb}\\
\lag_{D}&=&\frac{1}{8}\tr(\Phi^\dagger\Phi)
\tr[(D_\mu\Phi)^\dagger (D^\mu\Phi)].\label{ld}
\eea
\reseteqn
The term $\lag_W$
(quadrupole term)
contains nonstandard vector-boson self-interactions but
no interactions with scalar fields. \lwp\
and \lbp\ contain
vector-boson self-interactions and vector--scalar interactions.
\lww , \lbb\ and \ld\
contain terms quadratic in the gauge fields,
and thus it seems
as if they affect the vector-boson propagators. However, after
a redefiniton of the fields and coupling constants all
expressions that contain only gauge fields can be absorbed into the
SM Lagrangian \cite{ruetal}.
Thus these terms effectively only parametrize
nonstandard vector--scalar interactions.

In addition to these dimension-six terms we also consider
the following dimension eight-terms:
\alpheqn
\bea
\lwi&=&-\frac{1}{2}i\tr(\tau_3\Phi^\dagger W_\mu^{\,\,\,\nu}
W_\nu^{\,\,\,\lambda}\Phi)B_\lambda^{\,\,\,\mu}
,\label{lwi}\\
\lwpi&=&\frac{1}{4}i\tr[\tau_3(D_\mu\Phi)^\dagger
(D_\nu\Phi)]\tr(\tau_3\Phi^\dagger W^{\mu\nu}\Phi),\label{lwpi}\\
\lddi&=&-\frac{1}{4}\tr[(D_\mu\Phi)^\dagger(D_\nu\Phi)]
\tr[(D^\mu\Phi)^\dagger(D^\nu\Phi)],\label{lddi}\\
\lddii&=&-\frac{1}{4}\tr[(D_\mu\Phi)^\dagger(D^\mu\Phi)]
\tr[(D_\nu\Phi)^\dagger(D^\nu\Phi)].\label{lddii}
\eea
\reseteqn
All these terms contain vector-boson self-interactions
and vector--scalar interactions.
They are the phenomenologically
most interesting dimension-eight terms for the following reasons:
\lwi\ and \lwpi\ yield together with
the dimension-six terms \lw , \lwp\
and \lbp\ the most general, $C$, $P$, $CP$ and
locally \uem\ invariant trilinear vector-boson
self-interactions \cite{gore}. \lddi\ and
\lddii\ contain quadrilinear interactions
among the gauge fields but no
trilinear ones \cite{apbe,badava,gkkusc}.

Next we consider the case that the Higgs boson is very heavy or even
no Higgs boson exists. Then the effects of new physics should be
parametrized by taking the limit
$M_H\to \infty$ of an effective Lagrangian of type \eref{leffl}.
This limit is obtained by substituting the linearly realized scalar
fields in $\Phi$ \eref{phi} through a nonlinear expression
which contains only Goldstone bosons but no Higgs boson
\cite{apbe,gkko}:
\be
\Phi\to\frac{v}{\sqrt{2}}U
\label{linnonlin}
\ee
with
\be
U=\exp\left(\frac{i\vp_i\tau_i}{v}\right).
\label{u}
\ee
Applying \eref{linnonlin} to the SM \eref{lsm} one
obtains the gauged nonlinear $\sigma$ model
\cite{apbe} with the Lagrangian
\bea
\lag_{GNLSM}&=&-\frac{1}{2}\tr(W_{\mu\nu}W^{\mu\nu})-\frac{1}{4}
B_{\mu\nu}B^{\mu\nu}+\frac{1}{4}v^2\tr[(D_\mu U)^\dagger
(D^\mu U)]\nn&&+\mbox{fermionic terms},
\label{lgnlsm}
\eea
which is also
gauge invariant but nonrenormalizable.
Like in
the linear case, nonstandard interactions can be parametrized
by adding gauge invariant terms of higher dimension\footnote{Usually,
within nonlinear theories
the dimension is counted differently \cite{apbe,dohete,falusi,badava},
since $U$
\eref{u} has dimension 0, while the linear expression $\Phi$
\eref{phi}
has dimension 1. However, for consistency, we count the dimension
of nonlinear terms like in the linear case, thereby attributing the
dimension 1 to $U$, which is justified, since $U$ occurs
together with the vacuum expectation value
$v$ in \eref{linnonlin}. We will see that our
counting method makes sense when classifying the terms concerning the
high-energy behaviour they yield.}, in which the scalar sector is now
nonlinearly realized as \eref{u},
to the GNLSM Lagrangian.
The resulting Lagrangian
\be
\lag_{eff}^{NL}=\lag_{GNLSM}+
\sum_{n>4}\sum_i\frac{\ep_i}{\Lambda^{n-4}}
\lag_i^{(n)}
\label{leffnl}
\ee
is called a chiral Lagrangian.
Also the loop effects within the GNLSM generate such
effective interaction terms \cite{apbe}.

In this article we will consider the nonlinear
effective terms corresponding to
\eref{lw} -- \eref{lbp} and \eref{lwi} -- \eref{lddii}.
They are obtained from the
linear terms by the substitution \eref{linnonlin}.
Only \eref{lww} -- \eref{ld} can be neglected in
this case because by a field and coupling constant redefinition
they can be completely absorbed into the
GNLSM Lagrangian. We name the nonlinear terms like the
linear ones, since we will treat linear and nonlinear Lagrangians
seperately and no confusion can occur.

One could assume that a third possibility to construct effective
Lagrangians exists, namely to construct Lagrangians that are not gauge
invariant (except for the electromagnetic gauge freedom).
This possibility, however, is artificial because
each gauge noninvariant
Lagrangian can be written as a (nonlinear) gauge
theory such as \eref{leffnl}
by means of a Stueckelberg transformation
\cite{gkko,chgoge,bulo}. This means that
after applying a Stueckelberg
transformation in order to introduce unphysical scalar partners of
the massive vector fields, one can even apply the GET in order to
simplify calculations within originally
gauge noninvariant theories.

One can easily convince oneself that the GET \eref{get} is also
valid for effective gauge theories of the type \eref{leffl} or
\eref{leffnl}. In \cite{chga} only the invariance of the
quantized Lagrangian
under BRS transformations is used in order to prove the GET.
Renormalizability is not required in the derivation.
Since the additional
effective interaction terms are also gauge
invariant and thus BRS
invariant, they do not affect the validity of the
proof. However the explict form of
the BRS transformations is used in \cite{chga}. This is not changed
by adding effective interaction terms to the Lagrangian
but by parametrizing the scalar
sector nonlinearly as \eref{linnonlin}. Thus, the proof in
\cite{chga} only holds for linear effective Lagrangians.
An analogous
proof for the nonlinear case can be found in \cite{dope}. The general
validity of the GET
can most easily be seen from \cite{gokone}
where BRS invariance itself is used but not the explicit form of the
BRS transformations; thus the proof in \cite{gokone} applies to
linear and nonlinear effective gauge theories.

In order to apply power counting in the next section, we now
determine for each vertex deriving
from the each of the effective interaction terms
the power of the energy on which its
vertex factor depends. Obviously, this
power is identical to the number of derivatives in the
corresponding expression within the effective
term\footnote{Even within effective theories
the Feynman rules
can directly be obtained from the effective Lagrangian
in the usual way \cite{bedu}. This relies on the
equivalence of Hamiltonian and Lagrangian path integral
quantization.}, which is easily
found by substituting the definitions
\eref{not1} -- \eref{not2} and \eref{u} into the in the
effective term and then taking the traces.
We only consider
vertices that contribute --- after applying the GET \eref{get} ---
to the amplitudes for $VV\to VV$ and $f\bar{f}\to VV$
(in the tree approximation) .

Table \ref{vertices}
lists the (types of)
vertices that are of interest for us and
shows which vertices derive from each effective interaction
term and on which power of $E$ the corresponding
vertex factor depends.
The first part contains the SM vertices and the
additional contribution from the
linear effective interaction term; the second
part contains the vertices from
the GNLSM Lagrangian and from the corresponding nonlinear effective
terms. For better comparison, lines and columns
corresponding to interaction terms and to vertices that do not
occur in the nonlinear case are left free.
Terms that have a similar structure are put together in one column,
e.g. \lwp\ and \lbp\ are in one column denoted as
$\lag_{W\Phi,B\Phi}$, etc.
It should be noted that in general not all vertices of a
given type exist. For instance, $VVV$ vertices with three neutral
vector bosons do not exist, \lbp\ implies only vertices with at least
one
neutral vector boson, \ld , \lddi\ and \lddii\ yield no vertices
with photons, etc.

The effective terms yield
higher powers of $E$ in the vertices or even
induce vertices not present in the SM.
We will see
in the next section that, as one expects,
this will imply a worse high-energy behaviour of $S$-matrix
elements.


\section{The Power-Counting Method}
\eqnew
\typeout{Section 3}
In this section we will apply the GET \eref{get} in order to
analyze the high-energy behaviour of
scattering processes in the presence of
the effective interactions introduced above.
This will yield many explicit examples for the
invalidity of the ET within effective theories. We will show that
the GET simplifies calculations of
$S$-matrix elements because no cancellations occur.

In order to determine the high-energy behaviour
(i.e. the highest power of $E$ occurring in the amplitude)
of a process with
external longitudinally polarized vector bosons
within an effective theory and to
calculate the leading term of the
corresponding $S$-matrix element
(in the tree approximation) we proceed as
follows: We apply the GET
\eref{get} in order to express such an $S$-matrix element as
the sum of all amplitudes (calculated within the \rx -gauge)
in which each
longitudinal vector boson is
either replaced by the corresponding Goldstone boson
or its polarization vector by the nonleading term $v$ \eref{v}
(multiplied by a phase factor).
Then we construct all Feynman
diagrams that contribute to these amplitudes and
determine the high-energy behaviour
of each diagram.
We neglect all effects that are proportional to
$\ep_i\ep_j$ with $\ep_i$ and $\ep_j$ being effective coupling
constants associated with the higher dimension terms (see \eref{leffl}
and \eref{leffnl}), i.e. in all exchange
diagrams only one vertex is taken from an
effective interaction term but the other one from the SM or the
GNLSM,
respectively. The high-energy behaviour of the various diagrams can be
determined by simple power counting using
Table~\ref{vertices}. (An external $v$ line is $O(E^{-1})$ due to the
definition \eref{v}.) When applying
power counting within the linear models we always assume that
$E\gg M_H$;
the heavy-Higgs scenario is parametrized by the nonlinear models.

Tables~\ref{llll} -- \ref{tttt} and
\ref{ffll} -- \ref{fftt} display
the results of this power-counting
method applied to $VV\to VV$ and
$f\bar{f}\to VV$ for all effective interaction terms introduced in
Section~2. We have considered all
possible combinations of
polarizations of the vector bosons. (Processes with all $V$s being
transversal have only been considered for completeness.)
The following shorthand notation for the (types of)
Feynman diagrams was used:
The symbols outside the brackets denote the external lines and those
inside the brackets the exchanged particles such that
symbols written
on the same side of the bracket denote particles
coupled the same end of the propagator\footnote{
It should be noted, that e.g.\
$vv(V)\vp\vp$ corresponds to $vv\to \vp\vp$ or $\vp\vp\to vv$
if the vector boson is exchanged in the $s$-channel but
to $v\vp \to v\vp$ if the vector boson is exchanged in the
$t$- or $u$-channel; due to the phase factor in \eref{get}
both contributions have a relative factor $-1$.}.
$(C)$ stands for a
four-particle contact interaction.
The tables display in the first part the high-energy behaviour
of the various
Feynman diagrams within the SM and the additional
contributions of the linear effective interaction terms. The second
part shows the high-energy behaviour within the GNLSM and the
contributions of the corresponding nonlinear effective terms.

{}From these tables one can directly read off
the high-energy behaviour induced by a given
interaction term and
which diagrams contain the leading
contributions. Then one only has to calculate
these diagrams in order to obtain
the correct $S$-matrix element up to nonleading
terms. Thus, having once found the vertices implied
by the effective terms \eref{lw} -- \eref{ld} and
\eref{lwi} -- \eref{lddi}, our method enables one
to determine the high-energy behaviour of a process within a given
effective theory simply
by power counting without doing any
calculation, and it simplifies the calculation of the corresponding
amplitudes since only the leading parts of the
single diagrams have to be considered.
As mentioned in the introduction, power counting
cannot be applied if the GET
\eref{get} is not used and
diagrams with external $V_L$s have to be calculated, since
cancellations occur.

However, from the tables it is obvious that the ET \eref{et}
is not
valid for effective theories, because, in general, the leading
contributions of a given term do
not only (or not at all)
stem from diagrams in which all $V_L$s are
replaced by $\phi$s but also from diagrams with external $v$s.
In order to calculate these, one shold keep in mind that the
leading part of $v^\mu$ \eref{v} has the simple form
\be
v_\mu=-\frac{M}{2P_0^2}(P_0,-P_i)+O(E^{-3})
\label{vlead}.
\ee
If the in- or outgoing state consist of two particles
(1 and 2) with the same mass one finds in the CM system \cite{wi}
\be
v^\mu_1=-\frac{2M}{E^2}P^\mu_2+O(E^{-3}).
\ee
(with $E$ being the total energy) and vice versa for $v_2^\mu$.

In order to give an example how our method works,
let us look more closely
at the process $\vl\vl\to \vl \vl$ (Table~\ref{llll}).
By applying power counting to the diagrams that have to be
considered in a direct calculation without using
the GET \eref{get},
namely $\vl\vl\vl\vl (C)$, $\vl\vl(H,\vp)\vl\vl$ and
$\vl\vl(V)\vl\vl$,
one would expect that the amplitudes induced by the various terms
diverge as $O(E^4)$, the contributions of \lw\ and \lwi\ even
as $O(E^6)$. However, all amplitudes except for those stemming from
\lddi\ and \lddii\
behave at most as
$O(E^2)$, the effects of \lw , \lww , \lbb\ and \lwi\ are
even $O(E^0)$. Within a direct calculation this high-energy
behaviour is the result of cancellations; however it is obvious from
the GET \eref{get}.
Furthermore, Table~\ref{llll} shows that
for most of the effective terms the ET is not valid
because the leading part of the amplitude
gets contributions from diagrams with $v$s,
e.g. for \lw\ and \lwi\ it is given by the $vv(V)\vp\vp$ diagrams
while the diagrams with four external $\vp$s yield no contributions.
(For $\vt\vl\to \vl\vl$ (Table~\ref{tlll}),
the ET even would imply that
these two terms do not affect the good high-energy behaviour of the
SM; in fact they yield $O(E^{1})$ divergences.)

In may happen in some very
special cases that the high-energy behaviour
is even better than expected from the tables, because the Feynman
diagrams that would yield the leading contribution do not exist or
cancel accidentally.
We illustrate this by
considering the effects of \lw , \lwp\ and \lbp\ on
processes of the type $\vt\vt\to \vl\vl$. Table~\ref{ttll}
predicts that the contributions of these terms behave as $O(E^2)$
which is indeed the case for most of these processes \cite{ku}.
Let us however look closer at $W^+_TW^+_T\to W^+_LW^+_L$.
The $O(E^2)$
contribution of \lw\ would come from $W^+_TW^+_T(V)\vp^+\vp^+$
diagrams, which do not exist in this case.
For \lwp\ and \lbp\ also the $W^+_TW^+_T\vp^+\vp^+(C)$ and
$W^+_T\vp^+(H,\vp_0)W^+_T\vp^+$ diagrams should contribute.
However, it turns out that
\lbp\ affects none of these diagrams (it yields only vertices
with at least one neutral vector boson),
\lwp\ only contributes to $W^+_T\vp^+(H,\vp_0)W^+_T\vp^+$
and the $O(E^2)$ part of this contribution cancels.
Thus all three terms yield at most $O(E^0)$ corrections to
$W^+_TW^+_T\to W^+_LW^+_L$ \cite{ku}.
Another accidental cancellation is indicated in the
footnote of Table~\ref{tlll}.

The leading contributions of \lw , \lwp\ and \lbp\ (as long as they
are at least $O(E^1)$) on $VV\to VV$ were
calculated directly without applying the GET in \cite{ku}.
We have compared our results
with those of \cite{ku} and found that they agree.

The phenomenologically most important
results of our analysis are the following:
\begin{itemize}
\item All dimension-six terms \eref{lw} -- \eref{ld} yield
at most $O(E^2)$ contributions to the
$S$-matrix elements and thus, in the linear case, only
$O(E^0)$ contributions to the total cross-sections\footnote{
Remember that we neglect $\ep_i\ep_j$-terms.}
($\sigma \propto \M^2/E^2$).
I.e., they only slightly affect the good high-energy behaviour of the
SM. This result has been obtained for \lw , \lwp\ and \lbp\ in
\cite{ku}, we find it for all dimension-six terms.
\item The dimension-eight terms \eref{lwi} -- \eref{lddi} are
supposed to yield a worse high energy behaviour because of their
higher dimension. This is indeed the case for \lddi\ and \lddii\,
that imply $O(E^4)$ corrections to the $S$-matrix elements.
The terms \lwi\ and \lwpi\, however only yield $O(E^2)$
effects on the amplitudes\footnote{There is one exception, namely
\lwi\ yields an $O(E^3)$ effect on
$W^\pm_LW^\mp_T
Z^{\,}_T\gamma$ amplitudes (see Table~\ref{tttl}).
However, the amplitude in the SM
or in the GNLSM is $O(E^{-1})$ and thus
the contribution to the cross-section is only
$O(E^0)$.}, i.e. they behave in the same manner as the
dimension-six terms.
\item For most of the effective terms the worst high-energy behaviour
is associated with external longitudinal vector bosons, as one
expects. However the quadrupole terms \lw\ and \lwi\ behave well in
$\vl\vl\to \vl\vl$ and $f\bar{f}\to \vl\vl$ but yield increasing
amplitudes for $\vt\vt\to \vt\vt$ and $f\bar{f}\to \vt\vt$
\cite{gago,golare,ku}. Furthermore, the contribution of
\lwp , \lbp\ and \lwpi\
to $\vt\vt\vl\vl$-amplitudes is as divergent as the
one to $\vl\vl\vl\vl$-amplitudes, namely $O(E^2)$ \cite{ku}.
Thus
it is not sufficient to study only longitudinally polarized initial
and final states when looking for effects of new physics in
vector-boson scattering.
\item While in the SM the presence of a
(light) Higgs boson improves the high-energy behaviour (in the
(Higgs-less) GNLSM $\vl\vl\vl\vl$-amplitudes
are $O(E^2)$ but in the SM
they are $O(E^0)$
for $E\gg M_H$), this is not the case for the
higher-dimension extensions.
The high-energy behaviour of the effective
terms is the same in the linear case (with a light Higgs boson) and
in the nonlinear case (without a Higgs boson). Actually,
the value of the leading part of an $S$-matrix element
is affected by the presence of the
Higgs boson but the largest occurring power of $E$ is unchanged.
For \lw , \lwp\ and \lbp\ this has already been  found in \cite{ku}.
\end{itemize}

Our power-counting method is even useful within the SM
if one considers processes for
which the amplitude
is only $O(E^{-1})$ (see Tables~\ref{tlll}, \ref{tttl} and \ref{fftl}).
Although the ET
\eref{et} is formally correct in this case, it is obviously
of no use, because the corrections neglected in \eref{et} are
of the same order as the amplitude itsself. Applying
the GET \eref{get}
one can correctly determine the leading (i.e. $O(E^{-1})$)
part of the amplitude. Again this is not given by diagrams
with all $\vl$s being replaced by $\phi$s alone.

Although the ET
is in general not valid within effective theories,
there is an important exception.
Since the ET correctly determines the leading amplitudes
within the SM for $E\ll M_H$ \cite{ve}, it is also valid for the
leading parts of $S$-matrix elements calculated within the GNLSM
\footnote{Note, however, that
the relation
\eref{et} is {\em not\/} valid. The leading
(i.e.\ $O(E^2))$ tree-level contribution
to $\vl\vl\to
\vl\vl$ stems from the $\phi\phi\phi\phi(C)$ diagram and can
thus be determined by applying \eref{et}. However the corrections
from diagrams with $v$s are $O(E^0)$ (see Table~\ref{llll}) and not
only $O(E^{-1})$ as in \eref{et}.}
(without additional terms of higher dimension), which is
the heavy-Higgs limit of the SM.
Furthermore, the leading tree-level
contribution to the special process
$\vl\vl\to \vl\vl$ stems from the $\vp\vp\vp\vp(C)$ diagram
(see Table~\ref{llll}), i.e. it is
given by a pure scalar self interaction.
In \cite{veve} it has been shown that even the leading one-loop
corrections to this process
stem from diagrams in which all internal lines are scalar
lines.
Therefore, {\em in this special case\/}
even no internal vector lines have to be
considered and the leading amplitude can be obtained from a
Lagrangian which contains only scalar fields, namely the one
that is found by dropping all vector fields in the GNLSM Lagrangian
\eref{lgnlsm}:
\be
\lag_{scalar}=\frac{1}{4}v^2\tr[(\partial_\mu U)^\dagger
(\partial^\mu U)]
\label{lscalar}\ee
(nonlinear $\sigma$ model).
Since the GNLSM is the heavy-Higgs limit  of the SM,
the analogous
statement is true within the SM for $E\ll M_H$;
the leading tree-level
contribution to $\vl\vl\to \vl\vl$
comes from the $\phi\phi\phi\phi(C)$ and $\vp\vp(H)\vp\vp$
diagrams in this case\footnote{However,
note that for $E\gg M_H$ the vector exchange diagrams
$\vp\vp(V)\vp\vp$ must not be neglected.}
\cite{chga,ve}.


\section{Comparison with the Literature}
\typeout{Section 4}
\eqnew
After having shown how the equivalence theorem can be generalized such
that it even applies to effective Lagrangians we now critically
analyze those articles where the ET in the form \eref{et}, which
is in general not valid beyond the SM, was
used within effective theories.
We compare this treatment with the results of the
power-counting method described in the previous section
and examine whether the a priori incorrect use of \eref{et} led
to wrong results or not.

In \cite{chgoge,veve,ch} the ET was used within the GNLSM
in order to determine the amplitudes for
$\vl\vl\to \vl\vl$ in the tree approximation and at the one-loop
level.
Furthermore diagrams with internal vector lines
were neglected. As discussed in the previous section
this procedure is correct because no additional nonstandard interaction
terms were considered.
In \cite{caetal}
the same processes was investigated within the BESS model
(which is an extension of the GNLSM with one additional heavy
gauge-boson triplett) by applying the ET; this is also correct.

In \cite{dohete,falusi,badava}
the ET was applied to Lagrangians with additional
effective interaction terms of higher dimension.
In all these articles nonlinear effective Lagrangians of the type
\eref{leffnl},
i.e. Lagrangians without a Higgs boson were considered.
In \cite{falusi} the contributions of \lwp\
\eref{lwp} and \lbp\ \eref{lbp}
to the amplitude for $q\bar{q}\to VV$ were calculated by using
the ET. As one can see from Tables~\ref{ffll} -- \ref{fftt}, this is
correct; the leading contributions to the amplitudes are indeed found
by replacing all longitudinal vector bosons by Goldstone bosons.
In \cite{badava}, the amplitudes for $q\bar{q}\to \vl\vl$,
$\vl\vl\to \vl\vl$, $\vt\vl\to \vl\vl$ and $\vt\vt \to \vl\vl$,
were calculated within the GNLSM at the one-loop level and in
addition the tree-level
contributions of the (nonlinear versions of) the effective
terms \lwp\ \eref{lwp}, \lbp\ \eref{lbp}, \lddi \eref{lddi} and
\lddii\ \eref{lddii}  (and of two further effective terms that we
did not consider because they affect four-fermion amplitudes at the
tree level) were
determined by applying the ET. This procedure
is wrong because the ET is
not valid for the terms \lwp\ and \lbp .
Furthermore, it seems as if only contact diagrams were
calculated (see Figure~2 there).
To discuss this in more detail, let us consider the effects
of \lwp\ and \lbp\ on $Z_L Z_L\to W^+_LW^-_L$. In \cite{badava},
no $O(E^2)$-contributions of these terms were
found (see Appendix there), however such
contributions exist, as one can see from Table~\ref{llll}. (Also see
\cite{ku}.)
They stem from the $\vp\vp(V)\vp\vp$- and the $v\vp\vp\vp(C)$
diagrams. If one applies the ET \eref{et}, one neglects the
$v\vp\vp\vp(C)$ contribution and obtains a wrong result.
If one even considers only contact diagrams, one finds no
$O(E^2)$ effects of these terms at all, since they do not
affect the $\vp\vp\vp\vp(C)$ diagrams.

In \cite{dohete} the one-loop effects of the GNLSM and the
tree-level effects of effective higher-dimension terms
on $\vl\vl\to \vl\vl$ were
calculated, too.
In those articles not only the ET was used but also
internal vector boson lines were neglected, i.e.\ the GNLSM was
parametrized
by the pure scalar Lagrangian \eref{lscalar} and also in the
effective interaction terms all vector fields were set equal to zero.
As discussed in Section~3
this procedure is correct for the GNLSM.
However the treatment some
of the additinal effective interaction terms in
\cite{dohete} is wrong.
Terms like \lwp\ \eref{lwp}
and \lbp\ \eref{lbp}  were not considered
at all because they vanish for
vanishing gauge fields; their effects are simply
forgotten if one tries to describe effective interactions of vector bosons
by pure scalar self-interactions.

Thus we have found that in the articles, in which the ET
\eref{et} was
applied to effective Lagrangians, the
leading parts of $S$-matrix elements calculated
within the GNLSM
were correctly reproduced (for $\vl\vl\to \vl\vl$
even if internal vector lines were
neglected),
but incorrect results were obtained
for the contributions of several
additional effective interaction terms of higher dimension.

It should be noted, however, that
the numerical effects of this incorrect use of
the ET
are rather small as long as the coefficients of all effective
interaction terms are assumed to be of the same order of magnitude.
To explain this let us again consider
$\vl\vl \to \vl\vl$ (Table~\ref{llll}).
If one applys the ET, the contributions of
\lwp\ and \lbp\ are not correctly reproduced but
the correct contributions of
\lddi\ and \lddii\ are found.
The contributions of the
first two terms are $O(E^2)$, while those of the latter ones are
$O(E^4)$. Thus, if the coefficients of all
terms are of the same order of
magnitude, the correctly determined effects
of \lddi\ and \lddii\ are larger than
the incorrectly determined effects of
\lwp\ and \lbp . In $\vt\vt \to \vl\vl$ (Table~\ref{ttll})
all four terms yield $O(E^2)$ effects, but in this case the
ET is also valid for \lwp\ and \lbp .
However, if one assumes that the coefficients of
the dimension-eight
terms are much smaller than those of
the dimension-six terms, because they are
suppressed by higher negative powers of the scale of the
new physics $\Lambda$
\cite{lelora,ku,ruetal,gkkusc} (see \eref{leffl} and
\eref{leffnl}) the errors in these works become relevant and their
results cannot be applied.


\section{Summary}
\typeout{Section 5}
\eqnew
In this article we have pointed out that the equivalence theorem
cannot be applied to effective Lagrangians, which contain
nonstandard interaction terms of
higher dimension in addition to the Lagrangian of the
standard model or of the gauged nonlinear $\sigma$ model.

We have developed a correct method to simplify calculations of
$S$-matrix elements for scattering
processes with external longitudinally
polarized vector bosons within effective theories:
Instead of applying the equivalence theorem
\eref{et}, one has to use the  generalized equivalence theorem
\eref{get}, which also holds for
effective Lagrangians. On the basis of this theorem one can
determine the high-energy behaviour of $S$-matrix elements
simply by power counting without doing any calculations
and the corresponding amplitudes can be calculated more easily
since no cancellations occur.

We have applied this power-counting method to the phenomenologically
most important dimension-six and dimension-eight effective
interaction terms in  order to analyze their high-energy behaviour in
$VV\to VV$ and in $f\bar{f}\to VV$.
We have found that all dimension-six terms and even
some of the dimension-eight terms only slightly affect the good
high-energy behaviour
of the standard model, because their contributions to the
$S$-matrix elements are at most $O(E^2)$ and not $O(E^4)$ as one would
expect. Furthermore we have shown that the
addition of a light Higgs
boson does not improve the high-energy behaviour of these effective
terms.

Concerning the importance of our power-counting method we adopt a
statement made in \cite{wi} for the equivalence theorem
itself, namely that its true
value is not the simplification of calculations,
because everything can also be calculated directly without
applying the generalized equivalence theorem
\eref{get}
(as in \cite{ku}).
Its true value is that it gives a deep insight
in the qualitative nature of the effects of nonstandard interaction
terms on $S$-matrix elements, which is not directly obvious because
cancellations occur in a straightforward calculation.
Of course, if one calculates all $S$-matrix elements,
the high-energy behaviour of the contributions of the
various effective
terms can also be determined and the conclusions of the
previous paragraph can also be drawn,
however we have obtained these qualitative results
without calculations.

\section*{Acknowledgement}
We thank Reinhart K\"ogerler and Dieter Schildknecht for valuable
discussions.



\newpage


\section*{Tables}
\typeout{Tables}
\small
\begin{table}[h]
\begin{center}
\begin{tabular}{|l||c|c\sp c\sp c\sp c \sp
c\sp c\sp c|}
\hline
& \multicolumn{8}{|c|}{Linear Lagrangians}\\
\cline{2-9}
Vertex&\lsm&\lw&$\lag_{W\Phi,B\Phi}$&${\lag_{WW,BB}}^{\ss a}$&
${\lag_D}^{\ss a}$&\lwi&\lwpi&$\lag_{DD1,2}$\\ \hline
$\phi\phi\phi\phi$&0&--&--&--&2&--&--&4\\
$V\phi\phi\phi$&--&--&--&--&1&--&3&3\\
$VV\phi\phi$&0&--&2&2&0&--&2&2\\
$VVV\phi$&--&--&1&--&--&$3^{\ss b}$&1&1\\
$VVVV$&0&2&$0^{\ss c}$&--&--&2&0&0\\
$\phi\phi\phi$&--&--&--&--&--&--&--&--\\
$H\phi\phi$&0&--&--&--&2&--&--&--\\
$V\phi\phi$&1& --& 3&--&--&--&3&--\\
$VH\phi$&1& --& 3&--&$1$&--&3&--\\
$VV\vp$&0&--&2&--&--&--&2&--\\
$VVH$&0&--&2&$2$&$0$&--&2&--\\
$VVV$        & 1& 3& 1&--&--& 3& 1&--\\
$f\bar{f}V$&0&--&--&--&--&--&--&--\\ \hline\hline
& \multicolumn{8}{|c|}{Nonlinear Lagrangians}\\
\cline{2-9}
Vertex&\lgnlsm&\lw&$\lag_{W\Phi,B\Phi}$&--&
--&\lwi&\lwpi&$\lag_{DD1,2}$\\ \hline
$\phi\phi\phi\phi$&2&--&--&&&--&--&4\\
$V\phi\phi\phi$&1&--&3&&&--&3&3\\
$VV\phi\phi$&0&--&2&&&--&2&2\\
$VVV\phi$&--&--&1&&&$3^{\ss b}$&1&1\\
$VVVV$&0&2&$0^{\ss c}$&&&2&0&0\\
$\phi\phi\phi$&--&--&--&&&--&--&--\\
--&&&&&&&&\\
$V\phi\phi$&1& --& 3&&&--&3&--\\
--&&&&&&&&\\
$VV\phi$&0&--&2&&&--&2&--\\
--&&&&&&&&\\
$VVV$        & 1& 3& 1&&& 3& 1&--\\
$f\bar{f}V$&0&--&--&&&--&--&--\\ \hline
\multicolumn{9}{l}{\footnotesize$\quad{ }^a$After a field and
   coupling constant redefinition}\\
\multicolumn{9}{l}{\footnotesize$\quad{ }^b$Only $\gamma Z W^\pm
\phi^\mp$ vertices}\\
\multicolumn{9}{l}{\footnotesize$\quad{ }^c$Only \lwp}
\end{tabular}
\end{center}
\caption[]{\label{vertices}Vertices
deriving from the effective interaction terms, which contribute to
$VV\to VV$ and to $f\bar{f}\to VV$ after applying the GET
\eref{get}.
The integers denote the powers of the energy $E$ in the vertex
factors.}
\end{table}

\alphtab
\begin{table}[h]
\vspace*{-7mm}
\begin{center}
\begin{tabular}{|l||c|c\sp c\sp c\sp c \sp
c\sp c\sp c|}
\hline
\multicolumn{9}{|c|}{$V_LV_LV_LV_L$}\\
\hline
\hline
& \multicolumn{8}{|c|}{Linear Lagrangians}\\
\cline{2-9}
Diagram&\lsm&\lw&$\lag_{W\Phi,B\Phi}$&$\lag_{WW,BB}$&
$\lag_D$&\lwi&\lwpi&$\lag_{DD1,2}^\prime$\\ \hline
$\vp\vp\vp\vp(C)$&{\fb{0}}&--&--&--&{\fb{2}}&--&--&{\fb{4}}\\
$\vp\vp(H)\vp\vp$&\m2&--&--&--&0&--&--&--\\
$\vp\vp(V)\vp\vp$&{ \fb{0}}&--&{\fb{2}}&--&--&--&{\fb{2}}&--\\
\hline
$v\vp\vp\vp(C)$&--&--&--&--&0&--&{\fb{2}}&2\\
$v\vp(H)\vp\vp$&\m2&--&0&--&0&--&0&--\\
$v\vp(V)\vp\vp$&\m2&--&0&--&--&--&0&--\\
$vv\vp\vp(C)$&\m2&--&0&{\fb{0}}&\m2&--&0&0\\
$vv(H)\vp\vp$&\m4&--&\m2&\m2&\m2&--&\m2&--\\
$v\vp(H,\vp)v\vp$&\m2&--&0&--&\m2&--&0&--\\
$vv(V)\vp\vp$&\m2&{\fb{0}}&0&--&--&{\fb{0}}&0&--\\
$v\vp(V)v\vp$&\m4&--&\m2&--&--&--&\m2&--\\
$vvv\vp(C)$&--&--&\m2&--&--&--${}^{\ss a}$&\m2&\m2\\
$vv(H,\vp)v\vp$&\m4&--&\m2&\m2&\m4&--&\m2&--\\
$vv(V)v\vp$&\m4&\m2&\m2&--&--&\m2&\m2&--\\
$vvvv(C)$&\m4&\m2&\m4&--&--&\m2&\m4&\m4\\
$vv(H,\vp)vv$&\m6&--&\m4&\m4&\m6&--&\m4&--\\
$vv(V)vv$&\m4&\m2&\m4&--&--&\m2&\m4&--\\
\hline
&\yes&&\yes&&\yes&&&\yes\\
\hline\hline
& \multicolumn{8}{|c|}{Nonlinear Lagrangians}\\
\cline{2-9}
Diagram&\lgnlsm&\lw&$\lag_{W\Phi,B\Phi}$&--&
--&\lwi&\lwpi&$\lag_{DD1,2}^\prime$\\ \hline
$\vp\vp\vp\vp(C)$&{\fb{2}}&--&--&&&--&--&{\fb{4}}\\
--&&&&&&&&\\
$\vp\vp(V)\vp\vp$&{0}&--&{\fb{2}}&&&--&{\fb{2}}&--\\
\hline
$v\vp\vp\vp(C)$&0&--&{\fb{2}}&&&--&{\fb{2}}&2\\
--&&&&&&&&\\
$v\vp(V)\vp\vp$&\m2&--&0&&&--&0&--\\
$vv\vp\vp(C)$&\m2&--&0&&&--&0&0\\
--&&&&&&&&\\
$v\vp(\vp)v\vp$&\m2&--&0&&&--&0&--\\
$vv(V)\vp\vp$&\m2&{\fb{0}}&0&&&{\fb{0}}&0&--\\
$v\vp(V)v\vp$&\m4&--&\m2&&&--&\m2&--\\
$vvv\vp(C)$&--&--&\m2&&&--${}^{\ss a}$&\m2&\m2\\
$vv(\vp)v\vp$&\m4&--&\m2&&&--&\m2&--\\
$vv(V)v\vp$&\m4&\m2&\m2&&&\m2&\m2&--\\
$vvvv(C)$&\m4&\m2&\m4&&&\m2&\m4&\m4\\
$vv(\vp)vv$&\m6&--&\m4&&&--&\m4&--\\
$vv(V)vv$&\m4&\m2&\m4&&&\m2&\m4&--\\
\hline
&\yes&&&&&&&\yes\\
\hline
\multicolumn{9}{p{13cm}}{\footnotesize$\quad{ }^a$The $VVV\phi$ vertex
stemming from \lwi\ yields no contribution in this case, because one
$V$ is a photon, which cannot be longitudinal.}
\end{tabular}
\end{center}
\vspace*{-3mm}
\caption[]{\label{llll}The leading powers of the Energy $E$ in
the contribution (linear in the $\ep_i$)
of the effective interaction terms
to the
Feynman diagrams that yield $\M(V_LV_L\to V_LV_L)$ after
applying the GET
\eref{get}. The leading contribution of each
term is framed. The crosses denote the
terms for which the leading diagrams
are only those with all external $V_L$s being replaced by $\phi$s
(written above the horizontal line).
The shorthand notation for the (types of) Feynman diagrams is explained
in the text.}
\end{table}

\begin{table}[h]
\begin{center}
\begin{tabular}{|l||c|c\sp c\sp c\sp c \sp
c\sp c\sp c|}
\hline
\multicolumn{9}{|c|}{$V_TV_LV_LV_L$}\\
\hline
\hline
& \multicolumn{8}{|c|}{Linear Lagrangians}\\
\cline{2-9}
Diagram&\lsm&\lw&$\lag_{W\Phi,B\Phi}$&$\lag_{WW,BB}$&
$\lag_D$&\lwi&\lwpi&$\lag_{DD1,2}^\prime$\\ \hline
$\vt\vp\vp\vp(C)$&--&--&--&--&\fb{1}&--&\fb{$1^{\ss a}$}&\fb{3}\\
$\vt\vp(H)\vp\vp$&\fb{\m1}&--&\fb{1}&--&\fb{1}&--&\fb{1}&--\\
$\vt\vp(V)\vp\vp$&\fb{\m1}&--&\fb{1}&--&--&--&\fb{1}&--\\
\hline
$\vt v\vp\vp(C)$&\fb{\m1}&--&\fb{1}&\fb{1}&\m1&--&\fb{1}&1\\
$\vt v(H)\vp\vp$&\m3&--&\m1&\m1&\m1&--&\m1&--\\
$\vt\vp(H,\vp)v\vp$&\fb{\m1}&--&\fb{1}&--&\m1&--&\fb{1}&--\\
$\vt v(V)\vp\vp$&\fb{\m1}&\fb{1}&\fb{1}&--&--&\fb{1}&\fb{1}&--\\
$\vt\vp(V)v\vp$&\m3&--&\m1&--&--&--&\m1&--\\
$\vt vv\vp(C)$&--&--&\m1&--&--&\fb{1}&\m1&\m1\\
$\vt v(H,\vp)v\vp$, $v v(H,\vp)\vt\vp$&\m3&--&\m1&\m1&\m3&--&\m1&--\\
$\vt v(V)v\vp$, $vv(V)\vt\vp$&\m3&\m1&\m1&--&--&\m1&\m1&--\\
$\vt vvv(C)$&\m3&\m1&\m3&--&--&\m1&\m3&\m3\\
$\vt v(H,\vp)vv$&\m5&--&\m3&\m3&\m5&--&\m3&--\\
$\vt v(V)vv$&\m3&\m1&\m3&--&--&\m1&\m3&--\\
\hline
&&&&&\yes&&&\yes\\
\hline\hline
& \multicolumn{8}{|c|}{Nonlinear Lagrangians}\\
\cline{2-9}
Diagram&\lgnlsm&\lw&$\lag_{W\Phi,B\Phi}$&--&
--&\lwi&\lwpi&$\lag_{DD1,2}^\prime$\\ \hline
$\vt\vp\vp\vp(C)$&\fb{1}&--&
\fb{$1^{\ss a}$}&&&--&\fb{$1^{\ss a}$}&\fb{3}\\
--&&&&&&&&\\
$\vt\vp(V)\vp\vp$&{\m1}&--&\fb{1}&&&--&\fb{1}&--\\
\hline
$\vt v\vp\vp(C)$&{\m1}&--&\fb{1}&&&--&\fb{1}&1\\
--&&&&&&&&\\
$\vt\vp(\vp)v\vp$&{\m1}&--&\fb{1}&&&--&\fb{1}&--\\
$\vt v(V)\vp\vp$&{\m1}&\fb{1}&\fb{1}&&&\fb{1}&\fb{1}&--\\
$\vt\vp(V)v\vp$&\m3&--&\m1&&&--&\m1&--\\
$\vt vv\vp(C)$&--&--&\m1&&&\fb{1}&\m1&\m1\\
$\vt v(\vp)v\vp$, $v v(\vp)\vt\vp$&\m3&--&\m1&&&--&\m1&--\\
$\vt v(V)v\vp$, $vv(V)\vt\vp$&\m3&\m1&\m1&&&\m1&\m1&--\\
$\vt vvv(C)$&\m3&\m1&\m3&&&\m1&\m3&\m3\\
$\vt v(\vp)vv$&\m5&--&\m3&&&--&\m3&--\\
$\vt v(V)vv$&\m3&\m1&\m3&&&\m1&\m3&--\\
\hline
&\yes&&&&&&&\yes\\
\hline
\multicolumn{9}{l}{\footnotesize$\quad{ }^a$By power counting these
diagrams are $O(E^3)$, but all $O(E^3)$ terms cancel.}\\
\end{tabular}
\end{center}
\caption[]{\label{tlll}Same as Table~\ref{llll} for vector-boson
scattering with one transversal and three longitudinal
external
vector bosons: $\vt\vl\to \vl\vl$ and $\vl\vl\to \vt\vl$.}
\end{table}

\begin{table}[h]
\begin{center}
\begin{tabular}{|l||c|c\sp c\sp c\sp c \sp
c\sp c\sp c|}
\hline
\multicolumn{9}{|c|}{$V_TV_TV_LV_L$}\\
\hline
\hline
& \multicolumn{8}{|c|}{Linear Lagrangians}\\
\cline{2-9}
Diagram&\lsm&\lw&$\lag_{W\Phi,B\Phi}$&$\lag_{WW,BB}$&
$\lag_D$&\lwi&\lwpi&$\lag_{DD1,2}^\prime$\\ \hline
$\vt \vt \vp\vp(C)$&\fb{0}&--&\fb{2}&\fb{2}&\fb{0}&--&\fb{2}&\fb{2}\\
$\vt \vt(H)\vp\vp$&\m2&--&0&0&\fb{0}&--&0&--\\
$\vt\vp(H,\vp)\vt\vp$&\fb{0}&--&\fb{2}&--&--&--&\fb{2}&--\\
$\vt \vt(V)\vp\vp$&\fb{0}&\fb{2}&\fb{2}&--&--&\fb{2}&\fb{2}&--\\
$\vt\vp(V)\vt\vp$&\m2&--&0&--&--&--&0&--\\
\hline
$\vt \vt v\vp(C)$&--&--&0&--&--&\fb{2}&0&0\\
$\vt \vt(H,\vp)v\vp$, $\vt v(H,\vp)\vt\vp$&\m2&--&0&0&\m2&--&0&--\\
$\vt \vt(V)v\vp$, $\vt v(V)\vt\vp$&\m2&0&0&--&--&0&0&--\\
$\vt \vt vv(C)$&\m2&0&\m2&--&--&0&\m2&\m2\\
$\vt \vt(H,\vp)vv$, $\vt v(H,\vp)\vt v$&\m4&--&\m2&\m2&\m4&--&\m2&--\\
$\vt \vt(V)vv$, $\vt v(V)\vt v$&\m2&0&\m2&--&--&0&\m2&--\\
\hline
&\yes&\yes&\yes&\yes&\yes&&\yes&\yes\\
\hline\end{tabular}
\end{center}
\caption[]{\label{ttll}Same as Table~\ref{llll} for vector-boson
scattering with two transversal and two longitudinal
external
vector bosons: $\vt\vt\to \vl\vl$, $\vt\vl\to \vt\vl$ and
$\vl\vl\to \vt\vt$.
Only the linear effective terms are listed because
in the nonlinear case the corresponding powers of $E$ are the same.}
\end{table}

\begin{table}[h]
\begin{center}
\begin{tabular}{|l||c|c\sp c\sp c\sp c \sp
c\sp c\sp c|}
\hline
\multicolumn{9}{|c|}{$V_TV_TV_TV_L$}\\
\hline
\hline
Diagram&\lsm&\lw&$\lag_{W\Phi,B\Phi}$&$\lag_{WW,BB}$&
$\lag_D$&\lwi&\lwpi&$\lag_{DD1,2}^\prime$\\ \hline
$\vt \vt \vt \vp(C)$&--&--&\fb{1}&--&--&\fb{$3^{\ss a}$}&\fb{1}&\fb{1}\\
$\vt \vt(H,\vp)\vt\vp$&\fb{\m1}&--&\fb{1}&\fb{1}&\fb{\m1}&--&\fb{1}&--\\
$\vt \vt(V)\vt\vp$&\fb{\m1}&\fb{1}&\fb{1}&--&--&1&\fb{1}&--\\
\hline
$\vt \vt \vt v(C)$&\fb{\m1}&\fb{1}&\m1&--&--&1&\m1&\m1\\
$\vt \vt(H,\vp)\vt v$&\m3&--&\m1&\m1&\m3&--&\m1&--\\
$\vt \vt(V)\vt v$ &\fb{\m1}&\fb{1}&\m1&--&--&1&\m1&--\\
\hline
&&&\yes&\yes&\yes&\yes&\yes&\yes\\
\hline
\multicolumn{9}{l}{\footnotesize$\quad{ }^a$Only $W^\pm_L W^\mp_T
Z^{\,}_T \gamma$ reactions}
\end{tabular}
\end{center}
\caption[]{\label{tttl}Same as Table~\ref{ttll} for vector-boson
scattering with three transversal and one longitudinal
external
vector boson: $\vt\vt\to \vt\vl$ and $\vt\vl\to \vt\vt$.}
\end{table}

\begin{table}[h]
\begin{center}
\begin{tabular}{|l||c|c\sp c\sp c\sp c \sp
c\sp c\sp c|}
\hline
\multicolumn{9}{|c|}{$V_TV_TV_TV_L$}\\
\hline
\hline
Diagram&\lsm&\lw&$\lag_{W\Phi,B\Phi}$&$\lag_{WW,BB}$&
$\lag_D$&\lwi&\lwpi&$\lag_{DD1,2}^\prime$\\ \hline
$\vt \vt \vt \vt(C)$&\fb{0}&\fb{2}&\fb{0}&--&--&\fb{2}&\fb{0}&\fb{0}\\
$\vt \vt(H,\vp)\vt \vt$&\m2&--&\fb{0}&\fb{0}&\fb{\m2}&--&\fb{0}&--\\
$\vt \vt(V)\vt \vt$ &\fb{0}&\fb{2}&\fb{0}&--&--&\fb{2}&\fb{0}&--\\
\hline
\end{tabular}
\end{center}
\caption[]{\label{tttt}Same as Table~\ref{ttll} for
$\vt \vt \to \vt \vt$.}
\end{table}

\resettab
\alphtab

\begin{table}[h]
\begin{center}
\begin{tabular}{|l||c|c\sp c\sp c\sp c|}
\hline
\multicolumn{6}{|c|}{\rule{0mm}{4.5mm}$f\bar{f}\to \vl\vl$}\\
\hline
\hline
Diagram&\lsm&\lw&$\lag_{W\Phi,B\Phi}$&\lwi&\lwpi\\ \hline
$f\bar{f}(V)\vp\vp$&\fb{0}&--&\fb{2}&--&\fb{2}\\
\hline
$f\bar{f}(V)v \vp$&\m2&--&0&--&0\\
$f\bar{f}(V)vv$&\m2&\fb{0}&\m2&\fb{0}&\m2\\
$fv(f)\bar{f}v$&\m2&--&--&--&--\\
\hline
&\yes&&\yes&&\yes\\
\hline
\end{tabular}
\end{center}
\caption[]{\label{ffll}Same as Table~\ref{ttll} for
$f\bar{f}\to \vl\vl$. Diagrams with
$f\bar{f}H$ and $f\bar{f}\vp$ vertices
are not considered because they are negligible for
light fermions. The interaction
terms that are not listed yield no
contribution.}
\end{table}

\begin{table}[h]
\begin{center}
\begin{tabular}{|l||c|c\sp c\sp c\sp c|}
\hline
\multicolumn{6}{|c|}{\rule{0mm}{4.5mm}$f\bar{f}\to \vt\vl$}\\
\hline
\hline
Diagram&\lsm&\lw&$\lag_{W\Phi,B\Phi}$&\lwi&\lwpi\\ \hline
$f\bar{f}(V)\vt \vp$&\fb{\m1}&--&\fb{1}&--&\fb{1}\\
\hline
$f\bar{f}(V)\vt v$&\fb{\m1}&\fb{1}&\m1&\fb{1}&\m1\\
$f\vt(f)\bar{f}v$, $fv(f)\bar{f}\vt$&\fb{\m1}&--&--&--&--\\
\hline
&&&\yes&&\yes\\
\hline
\end{tabular}
\end{center}
\caption[]{\label{fftl}Same as Table~\ref{ffll} for
$f\bar{f}\to \vt\vl$.}
\end{table}

\begin{table}[h]
\begin{center}
\begin{tabular}{|l||c|c\sp c\sp c\sp c|}
\hline
\multicolumn{6}{|c|}{\rule{0mm}{4.5mm}$f\bar{f}\to \vt\vt$}\\
\hline
\hline
Diagram&\lsm&\lw&$\lag_{W\Phi,B\Phi}$&\lwi&\lwpi\\ \hline
$f\bar{f}(V)\vt \vt$&\fb{0}&\fb{2}&\fb{0}&\fb{2}&\fb{0}\\
$f\vt(f)\bar{f}\vt$&\fb{0}&--&--&--&--\\
\hline
\end{tabular}
\end{center}
\caption[]{\label{fftt}Same as Table~\ref{ffll} for
$f\bar{f}\to \vt\vt$.}
\end{table}

\resettab

\end{document}